\documentclass[aps,prl,twocolumn,groupedaddress]{revtex4-1}
\usepackage{graphicx}
\usepackage{amsmath}
\usepackage{dcolumn}
\usepackage{bm}

\begin{document}

\title{Spin coherence and dephasing of localized electrons in monolayer MoS$_2$}

\author{Luyi Yang$^1$, Weibing Chen$^2$, Kathleen M. McCreary$^3$, Berend T. Jonker$^3$, Jun Lou$^2$, Scott A. Crooker$^1$}

\affiliation{$^1$National High Magnetic Field Laboratory, Los Alamos, NM 87545, USA}
\affiliation{$^2$Department of Materials Science and NanoEngineering, Rice University, Houston, TX 77005, USA}
\affiliation{$^3$Materials Science and Technology Division, Naval Research Laboratory, Washington, DC 20375, USA}


\begin{abstract}
We report a systematic study of coherent spin precession and spin dephasing in electron-doped monolayer MoS$_2$. Using time-resolved Kerr rotation spectroscopy and applied in-plane magnetic fields, a nanosecond-timescale Larmor spin precession signal commensurate with $g$-factor $|g_0|\simeq 1.86$ is observed in several different MoS$_2$ samples grown by chemical vapor deposition.  The dephasing rate of this oscillatory signal increases linearly with magnetic field, suggesting that the coherence arises from a sub-ensemble of localized electron spins having an inhomogeneously-broadened distribution of $g$-factors, $g_0 + \Delta g$.  In contrast to $g_0$, $\Delta g$ is sample-dependent and ranges from 0.042-0.115. 
\end{abstract}
\maketitle

Atomically-thin crystals of the transition-metal dichalcogenides MoS$_2$, MoSe$_2$, WS$_2$, and WSe$_2$ are direct-bandgap semiconductors that, owing to their strong spin-orbit coupling and lack of inversion symmetry, exhibit novel spin- and valley-coupled physical properties such as valley-specific optical selection rules \cite{Xiao, MakNatNano, Zeng, Sallen, XuReview}. With a view towards exploiting both spin and valley degrees of freedom in these new 2D materials, early studies focused on the photoluminescence properties and fast picosecond dynamics of valley-polarized excitons \cite{Splendiani, Mak2010, Lagarde, Mai, Cui}.  More recent experiments, typically based on time- and polarization-resolved optical absorption or Faraday/Kerr effects, have begun to explore the intrinsic spin and valley dynamics of the background carriers (electrons or holes) that reside in $n$-type or $p$-type material \cite{Zhu, Plechinger, DalConte, Yang}.

One very recent study \cite{Yang} demonstrated a long-lived spin relaxation of resident electrons in monolayer MoS$_2$ crystals grown by chemical vapor deposition (CVD). In addition to the spin relaxation of mobile (itinerant) electrons due to the interplay between spin-orbit coupling and applied magnetic fields, a small but persistent signature of \emph{coherently-precessing} spins was also revealed.  Although speculated to arise from a sub-population of localized electrons, the coherence properties of these precessing spins were not, however, studied in detail. In particular, the relevant precession frequencies and decoherence timescales were not investigated systematically across a range of magnetic fields, temperatures, or samples grown under different conditions. As such, the origin of the spin coherence and its underlying dephasing mechanism has not been clearly established.

Here we directly address electron spin coherence in CVD-grown monolayer MoS$_2$. Time-resolved Kerr-rotation studies reveal clear signatures of Larmor spin precession persisting for many nanoseconds in several MoS$_2$ samples of different origin. The dephasing rate of these oscillatory signals is found to increase significantly and linearly with applied magnetic fields, strongly suggesting that the spin coherence arises from a sub-ensemble of independent and localized electron spins having an inhomogeneously-broadened distribution of Land\'{e} $g$-factors, $g_0 + \Delta g$.  While $|g_0| \simeq 1.86$ is observed in all samples studied, $\Delta g$ varies from sample to sample ($0.042 < \Delta g < 0.115$), likely reflecting different local disorder landscapes due to varying growth conditions.

\begin{figure}[b]
\center
\includegraphics[width=.4\textwidth]{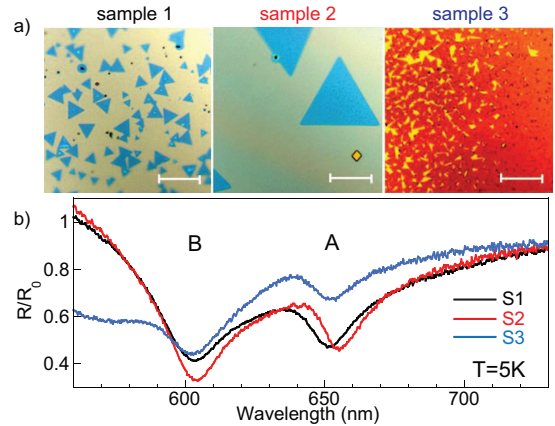}
\caption{(a) Images of the three CVD-grown monolayer MoS$_2$ samples studied in this work (S1, S2, S3). Scale bars: 50~$\mu$m. (b) Normalized reflectance spectra $R/R_0$ from all samples at 5~K. For S1 and S2, white light was focused to a small 10~$\mu$m spot within individual MoS$_2$ crystals. For S3, white light was focused on the continuous MoS$_2$ film. Clear A and B exciton features are observed for all samples.} \label{fig1}
\end{figure}

We investigated three different samples of high-quality monolayer MoS$_2$, all grown by CVD on SiO$_2$/Si substrates. Images of the samples are shown in Figure 1(a). Samples 1 and 2 (S1 and S2; both grown at Rice University) contain isolated monolayer MoS$_2$ crystals \cite{Lou} having typical lateral dimensions of $\sim15~\mu$m and 75~$\mu$m, respectively. MoO$_3$ and pure sulfur powder were used as precursor and reactant materials, and the growth was performed at a reactant temperature of $\sim$750 $^\circ$C on blank SiO$_2$/Si substrates. S1 was grown at slightly lower temperature and used more precursor than S2, leading to easier nucleation and a higher density of smaller MoS$_2$ crystals. Sample 3 (S3; grown at the Naval Research Lab) used the same precursor and reactant materials, but was grown using 625 $^\circ$C reactant temperature. In addition, perylene-3,4,9,10-tetracarboxylic acid tetrapotassium salt was loaded on the SiO$_2$/Si substrate, which acted as a seeding promoter to achieve uniform large-area monolayer MoS$_2$ \cite{Ling}. S3 has regions of both discrete and continuous monolayer MoS$_2$.

Individual MoS$_2$ crystals within S1 and S2 (and regions of MoS$_2$ within S3) were screened for good optical quality at low temperatures based on reflectance spectroscopy and also on a high degree of circularly-polarized photoluminescence ($>$75\%) when exciting with circularly-polarized light at 632.8~nm.  Figure 1(b) shows characteristic reflectivity spectra from all three samples at 5~K.  Both the A and B exciton features are clearly resolved in all samples. All the MoS$_2$ samples have residual $n$-type (electron) doping.  In S1 and S2, the background electron density $n_e$ is estimated to be on the order of $5\times10^{12}$ cm$^{-2}$ based on transport studies \cite{Lou} of field-effect transistors fabricated from similarly-grown monolayer MoS$_2$.

To directly measure the spin dynamics of these resident electrons, we use ultrafast optical techniques based on time-resolved Kerr rotation (TRKR). In contrast to photoluminescence (PL) measurements -- which necessarily require the participation of a photo-excited hole and which therefore primarily reveal \emph{exciton} dynamics -- TRKR can directly detect both exciton dynamics as well as the coupled spin/valley polarization dynamics of the background resident electrons in the conduction band of \emph{n}-type MoS$_2$. Crucially, these intrinsic electron dynamics can persist long after all the holes have recombined and PL has ceased.

Figure 2(a) depicts the experimental setup. The samples were mounted in vacuum on the cold finger of a small optical cryostat (3-300~K) that is affixed to an $xyz$ positioning stage. External coils can apply transverse magnetic fields $B_y$. The TRKR experiments used wavelength-degenerate 250~fs pump and probe pulses from a 76~MHz optical parametric oscillator, typically tuned to the low-energy side of the fundamental A exciton resonance in MoS$_2$ ($\sim$660~nm). The pump beam was right- or left-circularly polarized (RCP or LCP) to excite spin- (and valley-) polarized electrons and holes initially oriented along $\pm \hat{z}$, respectively, at $t$=0. To facilitate lock-in detection, the pump beam was either intensity-modulated by a chopper, or was polarization-modulated between RCP and LCP by a photoelastic modulator. To mitigate any influence of carrier diffusion or density gradients in our measurements, the pump beam (200~$\mu$W average power) was weakly focused to uniformly illuminate a large 25~$\mu$m spot, while the probe beam (tens of $\mu$W) was more tightly focused to a 4~$\mu$m spot positioned in the center of the triangular crystals.

\begin{figure}[tbp]
\center
\includegraphics[width=.42\textwidth]{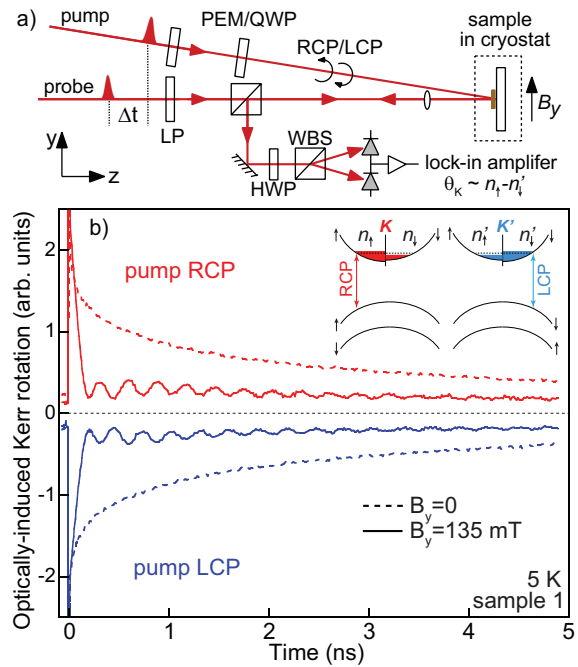}
\caption{(a) Schematic of the time-resolved Kerr rotation (TRKR) experiment.  Spin- and valley-polarized electrons (and holes) are resonantly photoexcited into monolayer MoS$_2$ at the A exciton using right- or left-circularly polarized (RCP/LCP) pump pulses. The induced spin/valley polarization along $\hat{z}$ is detected via the Kerr rotation $\theta_K$ imparted on linearly-polarized and time-delayed probe pulses. LP: linear polarizer, PEM: photoelastic modulator, QWP: quarter-wave plate, HWP: half-wave plate, WBS: Wollaston beam splitter. (b) Induced $\theta_K$ signals from MoS$_2$ at 5~K for both RCP and LCP pump, in transverse fields $B_y$=0 and 135~mT. Peak Kerr rotation angles at $t \sim 0$ are approximately 600 microradians. The signal inverts sign when reversing the pump handedness (as expected), and exhibits small but clear oscillations when $|B_y|>0$, indicating coherent spin precession of some (localized) electrons. Inset: a single-electron picture of the conduction and valence bands at the $K$ and $K'$ valleys of monolayer MoS$_2$, and the relevant spin- and valley-specific optical selection rules. For clarity, the spin-up and spin-down conduction bands are separately drawn on the left and right side within each valley, respectively.} \label{fig2}
\end{figure}

The pump-induced spin/valley polarization of the resident electrons along the $\hat{z}$ direction, $s_z(t)$, was subsequently detected via the Kerr rotation $\theta_K$ imparted on the time-delayed and linearly-polarized probe pulses that were reflected at normal incidence from the MoS$_2$. $\theta_K$ is proportional to the \emph{difference} between the RCP and LCP absorption constants and refraction indices of MoS$_2$. For wavelengths near the fundamental A exciton transition, these RCP and LCP optical properties depend sensitively on the density of spin-up and spin-down electrons residing in the $K$ and $K'$ valley ($n_\uparrow$ and $n'_\downarrow$, respectively; see inset, Fig. 2b). This is due to the optical selection rules in monolayer MoS$_2$ which mandate \cite{Xiao} that near the A exciton transition, RCP light couples primarily to spin-up electrons in the $K$ valley, while LCP light couples primarily to spin-down electrons in the $K'$ valley. As such, $\theta_K \propto n_\uparrow -n'_\downarrow$, and contains contributions from both electron spin and valley polarization. We note that similar time-resolved Kerr/Faraday rotation methods have been extensively applied to measure the spin dynamics of resident background electrons in quantum wells and quantum dots in conventional semiconductors (like GaAs and CdTe) that possess related (spin-selective) optical selection rules \cite{Awsch, Zhukov, Chen, Greilich}. A key distinction is that the fundamental exciton and trion optical resonances in MoS$_2$ occur at the $K/K'$ points of the Brillouin zone (as opposed to the $\Gamma$ point), and that the selection rules in MoS$_2$ allow not only spin but also valley selectivity.

Figure 2(b) shows the pump-induced $\theta_K$ from MoS$_2$ at 5~K for both RCP and LCP pump light. Dashed and solid lines correspond to $B_y$=0 and 135~mT, respectively.  As expected for this polarization-sensitive measurement, the signal inverts sign when reversing the handedness of the pump. Consistent with recent studies of similar MoS$_2$ samples \cite{Yang}, $\theta_K$  decays slowly on $\sim$3~ns timescales at zero field, which exceeds reported PL recombination times by 2-3 orders of magnitude \cite{Lagarde}. This indicates that the injected carriers have `imprinted' a nonequilibrium polarization on the sea of resident electrons, which subsequently relaxes slowly.  At $B_y$=135~mT, however, most of the TRKR signal quickly decays within $\sim$100~ps. As described previously \cite{Yang}, this is believed to arise from the rapid depolarization of the itinerant electron spins caused by their precession about the combined applied field $B_y \hat{y}$ and the large effective spin-orbit field $\pm B_{\rm{SO}}\hat{z}$ that is `seen' by mobile electrons in the $K$/$K'$ valley, which fluctuates rapidly due to fast inter-valley electron scattering. Crucially, however, a small but surprisingly long-lived oscillatory signal remains behind and persists for several nanoseconds, indicating that \emph{some} electrons undergo coherent spin precession. This signal, which is not expected from itinerant resident electrons for the reasons just described, likely arises from localized or trapped electron states that are not subject to rapid intervalley scattering and which precess only about the bare applied field $B_y$. The detailed dynamics and dephasing properties of these coherently-precessing spins, which has not been studied to date, is the primary focus of this work.

\begin{figure}[tbp]
\center
\includegraphics[width=.42\textwidth]{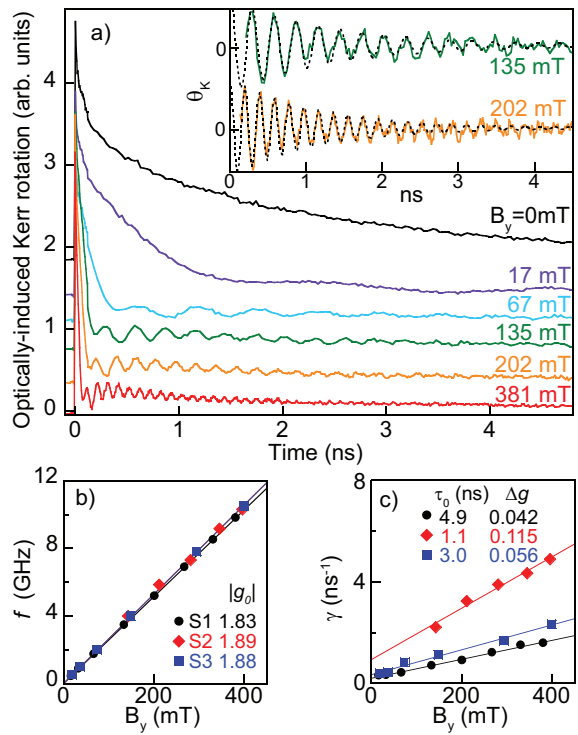}
\caption{(a) Induced $\theta_K$ from MoS$_2$ (sample 1) at 5~K for different $B_y$. Curves offset for clarity. Inset: expanded view of the oscillatory spin coherence signal at $B_y$=135 and 202~mT (a smoothly-varying background was subtracted). The precession signals can be well fit by an exponentially-decaying cosine, $Ae^{-\gamma t}\textrm{cos}(2\pi f t)$ (dashed lines). (b) The measured precession frequency $f$ increases linearly with $B_y$, indicating a similar average $g$-factor ($|g_0| \simeq$1.86) for all MoS$_2$ samples studied. (c) The measured decay rate ($\gamma$) also increases linearly with $B_y$, indicating an inhomogeneous distribution of $g$-factors ($g_0 + \Delta g$) in all samples, consistent with a sub-ensemble of independent, localized electron spins. The lines are linear fits to $\gamma = 1/\tau_0 + \Delta g \mu_B B_y/\hbar$.} \label{fig3}
\end{figure}

Figure 3(a) shows TRKR data for increasing $B_y$, where the appearance and evolution of the coherent signal can be clearly observed. Both the frequency and the decay rate of the precession signals increase with $B_y$.  Following the subtraction of a smoothly-varying background, the inset of Fig. 3(a) shows that these remaining oscillatory signals can be fit quite well to an exponentially-decaying cosine, $s_z(t)=A e^{-\gamma t}\textrm{cos}(2\pi f t)$. Note, therefore, that these signals commence with non-zero amplitude at $t=0$, as expected for an electron spin polarization that is quickly `initialized' along $\hat{z}$ by the photoinjection of spin-polarized electrons, but in contrast to the sinusoidal spin precession that is typical of embedded magnetic impurities (for example, in diluted magnetic semiconductors \cite{CrookerPRB}). To gain insight into this long-lived spin coherence and to elucidate the physical mechanism underpinning its decay, we fit the TRKR data from all three MoS$_2$ samples, and plot the precession frequency $f$ and decay rate $\gamma$ versus $B_y$ in Figs. 3(b) and 3(c). Two noteworthy features are apparent:

1) In all samples, $f$ increases linearly with $B_y$, as expected for simple Larmor spin precession about $B_y$ alone; \emph{viz.} $2\pi f = g_0 \mu_B B_y/\hbar$, where $\mu_B$ is the Bohr magneton. Moreover, the slope of $f(B_y)$ is essentially independent of sample, indicating nearly identical Land\'{e} $g$-factors  in all the MoS$_2$ samples studied ($|g_0| \simeq 1.86$).

2) More importantly, $\gamma$ also increases markedly and linearly with $B_y$ in all samples. This strongly suggests that the decay is dominated by ensemble spin dephasing due to an inhomogeneously-broadened distribution of electron $g$-factors, $g_0 + \Delta g$, within a given MoS$_2$ crystal.

A collection of independent spins with a distribution of $g$-factors $g_0 + \Delta g$ (and therefore a range of precession frequencies $f+\Delta f$) will exhibit an ensemble dephasing rate that increases linearly with $B_y$; namely $\gamma = 1/\tau_0 + \Delta g \mu_B B_y/\hbar$, where $\tau_0$ is the intrinsic decoherence time at zero field.  This behavior has been observed, \emph{e.g.}, in spin dephasing studies of epitaxial and colloidal quantum dot ensembles \cite{Greilich, Gupta, Berezovsky}, where the exact $g$-factor of a particular confined electron spin depends on its local environment and confinement potential, which varies from dot to dot.  Note that this discussion of exponential decays presupposes a Lorentzian distribution for $\Delta g$. While this is a reasonable approximation in that it captures the observed trends [see inset, Fig. 3(a)], the data do not have sufficient signal-to-noise to reliably distinguish the small differences in decay dynamics that arise from other (\emph{e.g.}, Gaussian) functional forms.

Interestingly, the very different slopes of the three $\gamma(B_y)$ traces in Fig. 3(c) indicates that the inhomogeneous broadening $\Delta g$ varies significantly from sample to sample ($\Delta g$ ranges from $\sim$0.04 to 0.12), suggesting that the local disorder landscape in CVD-grown MoS$_2$ is sensitive to the details of sample growth and processing. In addition, the intrinsic decoherence time of the precessing electron ensemble, $\tau_0$, also varies significantly from 4.9~ns down to 1.1~ns, with shorter $\tau_0$ correlated with larger $\Delta g$.

\begin{figure}[tbp]
\center
\includegraphics[width=.42\textwidth]{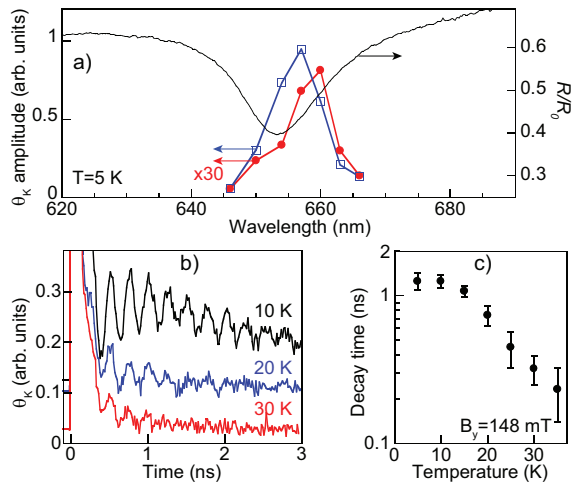}
\caption{(a) The measured amplitude of the oscillatory signal vs. laser wavelength (red dots). The blue squares show the initial amplitude of the TRKR signal at $t\sim$0. Also shown is the normalized reflectivity $R/R_0$ from the sample, showing the A exciton resonance. (b) TRKR of MoS$_2$ (sample 1) at 10, 20, and 30~K. $B_y$=148~mT; curves offset for clarity. (c) The decay time of the precession signal vs. temperature. Error bars show the $\chi^2$ uncertainty of a damped-cosine fit.} \label{fig4}
\end{figure}

The amplitude of the coherence signal is small, and does vary from sample to sample (in S1, S2, and S3 the amplitude is 25, 40, and 10 times smaller, respectively, than the initial $\theta_K$ measured at $t \sim$0). Its dependence on the pump/probe laser wavelength is shown in Fig. 4(a). While the precession frequency $f$ remains unchanged (not shown), the amplitude of the precession signal is largest when the laser is tuned to energies just below the peak of the A exciton resonance, again consistent with localized electron states in these MoS$_2$ monolayers. Localized `quantum-dot-like' emitting states below the A exciton resonance have recently been studied by several groups in monolayer WSe$_2$ \cite{Srivastava, He, Koperski, Chakraborty}.

Finally, the temperature dependence of the spin coherence is shown in Figs. 4(b,c). While $f$ remains unchanged upon increasing temperature (to within fitting error), the decay rate $\gamma$ accelerates significantly and oscillatory signals are not observed above 35~K. This may reflect a characteristic energy scale of a few meV for thermal excitation out of localized states, or for phonon- or spin-orbit-induced spin relaxation mechanisms in these monolayer materials that have been studied in recent theoretical works \cite{Dery, Ochoa, Wu, Glazov}.

In summary, time-resolved Kerr rotation studies of CVD-grown monolayer MoS$_2$ samples reveal a long-lived oscillatory signal persisting for nanosecond timescales at low temperatures.  These signals are consistent with coherent spin precession of a sub-ensemble of localized electrons in these $n$-type samples. The field dependence of the measured dephasing rate suggest that these localized electrons possess an inhomogeneously-broadened distribution of $g$-factors, $g_0 + \Delta g$. While $|g_0| \simeq 1.86$ is approximately the same in all samples measured, $\Delta g$ varies considerably from sample to sample, likely due to growth- and processing-dependent variations in the local disorder landscape.  Although presently observed only below 30~K, the robust spin coherence in monolayer MoS$_2$ may find applications in spin-based devices in these new monolayer dichalcogenides.

This work was supported by the Los Alamos LDRD program. These optical studies were performed at the National High Magnetic Field Laboratory, which is supported by NSF DMR-1157490 and the State of Florida. We also acknowledge the support from AFOSR (grant FA9550-14-1-0268) and the Welch Foundation (grant C1716).  The work at NRL was supported by core programs and the NRL Nanoscience Institute, and by the Air Force Office of Scientific Research under contract number AOARD 14IOA018-134141.

\end{document}